\documentclass[12pt,preprint]{emulateapj}
\usepackage[]{natbib}

\usepackage{graphicx}    

\newcommand\msun{\rm M_{\odot}}

\newcommand\msunyr{\rm M_{\odot}\,yr^{-1}}
\newcommand\be{\begin{equation}}
\newcommand\en{\end{equation}}

\shorttitle{The Extreme Outflow GGD 37}
\shortauthors{Green et al.}

\begin{document}

\title{GGD 37: An Extreme Protostellar Outflow}

\author{J. D. Green\altaffilmark{1},
 D. M. Watson\altaffilmark{2}, E. Bergin\altaffilmark{3}, S. Maret\altaffilmark{4}, G. Melnick\altaffilmark{5}, P. Sonnentrucker\altaffilmark{6}, V. Tolls\altaffilmark{5}, B. A. Sargent\altaffilmark{6}, W. J. Forrest\altaffilmark{2}, K.H. Kim\altaffilmark{2}, \& S. N. Raines\altaffilmark{7}}

\affil{1. Department of Astronomy, University of Texas at Austin;
2. Department of Physics and Astronomy,
University of Rochester, Rochester, NY 14627;
3. Department of Astronomy, University of Michigan;
4. Laboratoire d'Astrophysique de Grenoble;
5. Harvard-Smithsonian Center for Astrophysics, 60 Garden St., Cambridge, MA 02138;
6. Space Telescope Science Institute, Baltimore, MD;
7. Department of Astronomy, University of Florida, Gainesville, FL}

\begin{abstract}
We present the first Spitzer-IRS spectral maps of the Herbig-Haro flow GGD 37 detected in lines of [Ne III], [O IV], [Ar III], and [Ne V].  The detection of extended [O IV] (55 eV) and some extended emission in [Ne V] (97 eV) indicates a shock temperature in excess of 100,000 K, in agreement with X-ray observations, and a shock speed in excess of 200 km s$^{-1}$.  The presence of an extended photoionization or collisional ionization region indicates that GGD 37 is a highly unusual protostellar outflow.
\end{abstract}

\section{Introduction}

Protostellar cores of all masses eject material in the form of detectable bipolar outflows when their accretion rates are sufficiently high, ie. $\gtrsim$ 10$^{-7}$ $\msunyr$.  Very massive stars may trigger additional star formation through these powerful flows.  Even the much more numerous outflows from low mass stars may impact their local environment with high velocity flows and stir up turbulence.  Intermediate mass stars are an interesting middle scenario: are they numerous enough and powerful enough to trigger further star formation, or suppress it?  One way to determine their effect on their environment is to analyze the physical conditions of the surrounding material before and after it is impacted by shocks tied to the jet/outflow.

Shocks were categorized by their physics in \citet{draine80}: J-type shocks in which the magnetic field from the shocked ions is either frozen into the shock, or is non-existent; J-type shocks with magnetic precursors (or radiative precursors), in which the neutral fluid undergoes a discontinuous ``jump'' in density and temperature; and C-type shocks in which the fluid changes continuously over the shock boundary.  The conditions in these shocks vary and can be constrained through observations of fine structure line emission with differing appearance potential and critical density.  At 700 pc distant, the Cepheus A cloud complex contains a number of Herbig-Haro objects -- natural laboratories for shock models -- housed within the Cep OB3 association.

\begin{figure*}
\begin{center}
\includegraphics[width=0.45\textwidth]{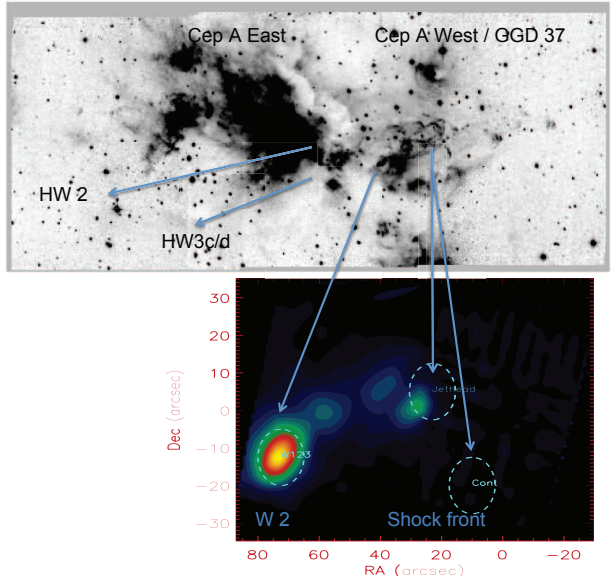}
\caption{Top: near-IR (K'-band) image of GGD 37 \citep{hodapp94} in continuum and H$_2$, with key positions annotated.  Bottom: Location of the 15'' HPBW apertures used to extract individual spectra in GGD 37, overplotted on the [O IV] line map, with corresponding positions in the near-IR map.}
\label{apertures}
\end{center}
\end{figure*}

Cep A contains a collection of luminous protostellar sources; additionally these protostars are associated with at least two broadly extended CO outflows, Cep A East and Cep A West \citep{hughes82}; the latter sometimes referred to as GGD 37 \citep{gyulbudaghian78} or HH 168 \citep{reipurth99}.  Both flows have been mapped by Spitzer-IRS \citep{sonnentrucker06,neufeld06b}. The extinction toward GGD 37 is generally thought to be small \citep{froebrich02} but \citet{wright96} apply an A$_V$ of 16 to their data.  High-resolution near-IR imaging \citep{raines00b} of the molecular emission from GGD 37 reveals a chain of bow shocks in GGD37.  The SE-NW orientation of most of these shocks suggest an outflow driven from the vicinity of HW3c (a 10 $\msun$ protostar in Cep A East), while other bow shocks suggest a flow orientation that is more east-west.  Additionally a chain of several extremely high proper motion ($\sim$ 800 km s$^{-1}$) knots of [Fe II] {1.644 $\mu$m} emission connects the SE-NW flow but is notably absent in the E-W flow.  The [Fe II] emission peaks around a trio of radio sources, centered at ``W2'' \citep{garay96} (see Figure \ref{apertures}).

The Cep A region has been mapped in both hard and soft X-rays with XMM-Newton \citep{pravdo05,pravdo09} and Chandra \citep{schneider09a}.  Both HW2 and HW3c are detected in hard X-rays, while GGD 37 (including the W1-W2-W3 region) is detected in soft X-rays.  \citet{pravdo05} note several distinct soft X-ray sources within the GGD 37 flow, one near the W2 region, and one at the far end of the flow.  They conclude for GGD 37 that L$_X = 3 \times 10^{30}$ erg s$^{-1}$, and infer a shock velocity of 620 km s$^{-1}$, and a temperature of 4 $\times$ 10$^6$ K from comparisons to L1551-IRS 5 / HH 154 \citep{favata02,bally03}. They observe that the flow is extreme in temperature, although not in luminosity.  The X-ray emissions are offset from both the optical and radio emissions, and they conclude this is due to the complex morphology of the region and multiple driving sources, as well as anisotropy in the medium.  In this paper we present Spitzer-IRS maps of highly ionized mid-IR forbidden lines in GGD 37 and show that the offset and extended nature of these {\it Spitzer} lines, and the X-ray emission, are directly attributable to the cloud shock.

\section{Observations}

\subsection{Data Reduction}

We utilized the InfraRed Spectrograph (IRS) onboard {\it Spitzer} to observe GGD 37 as part of our guaranteed and general observing time (program 2, PI: J.R. Houck; program 30167, PI: D. A. Neufeld); for the full dataset, see Green et al. (2010a, in prep.).  We observed this region in a Spitzer-IRS map grid consisting of two adjacent rectangles, on two separate occasions; first in Short-High (SH; spectral resolution R $\sim$ 600) and Long-High (LH; R $\sim$ 600) in 2004, and again in Short-Low (SL; R $\sim$ 60-120) and in deeper exposures in Long-High in 2007.  These maps cover approximately the same regions of the sky, although Cep A East was observed concurrently with the first set of observations of GGD 37.  The resolution of individual pixels of the IRS is given in the Spitzer Observers' Manual \footnote{http://ssc.spitzer.caltech.edu/documents/som/} as ranging from $\sim$ 2 - 5 arcsec squares, depending upon the module in question; therefore the maps contain differing spatial resolution, although they have all been regridded and oversampled for display purposes.

For full details of the data reduction process, see Green et al. (2010a); here we briefly summarize the procedure.  The basic calibrated data (BCD) were processed at the Spitzer Science Center (SSC) using version S12.0, S15.3, or S17.2 of the processing pipeline, and then reduced using SMART \citep{higdon04}, modified by additional routines that we have developed to deal with map grids (as opposed to single observation staring mode).  We process the data beyond the traditional SMART reduction by removing bad pixels in all modules, using a ``grand" rogue mask created from a superposition of the bad pixel masks of each separate observing campaign.  Rogues that occur once at the 4$\sigma$ level are considered to be permanently bad pixels and thus the grand rogue mask contains 25\% more rogue pixels than a single campaign rogue mask.   Bad pixels are fixed using a nearest neighbor routine (imclean.pro and imnan.pro).

Next we extracted individual spectra not simply from each slit position, but from each individual resolution element of each slit position.  In the case of SH and LH, we produce five (semi-)independent spectra, spatially separated by the resolution of a single pixel in the array; in SL we use 32 independent positions to generate spectra.  The spectra are then re-gridded onto a regular grid in right ascension and declination.  Pixels with (nearly) identical positions are averaged together.  We apply the slit loss correction function (J. D. Smith, private communication) to correct for flux-calibration created for point-sources, used by the SSC.  The result is a downward correction to the flux to $\sim$ 60-90\% of the original value.

\subsection{Extended Detection of Lines with Appearance Potential $>$ 13.6 eV}

We report the first detection of [Ar III], [Ne III], [O IV] or [Ne V] in a Herbig-Haro flow.  The [O IV] (25.89 $\mu$m) detection is fit separately both spectrally and spatially from the much brighter nearby [Fe II] line (25.99 $\mu$m).    \citet{guiles07} observed emission from apparent spectral lines several resolution elements on both sides of extremely bright lines in their sources, but they were ascribed to data artifacts due to the symmetric nature of the ``lines" around the central peak.  In this case the [Fe II] line brightness is not extreme and there is no corresponding bump on the long-wavelength side of the [Fe II] 26 $\mu$m line.  The [Ar III] map is co-spatial with the other lines, with the exception that it appears to contribute to the westernmost clump in the map, unlike the other species of higher appearance potential.  Additionally there is a possible detection of [Ar III] (8.99 $\mu$m) but it is suspect due to lack of coherence in its mapped spatial structure (Green et al. 2010b, in prep.).

\begin{figure}
\begin{center}
\includegraphics[scale=0.45]{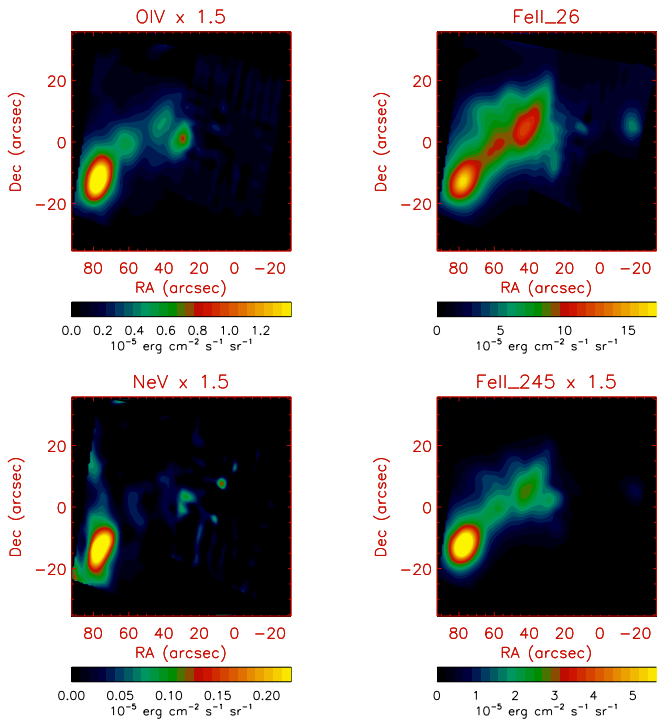}
\caption{Top Left: Map of [O IV] emission from W2 (lower left corner) to the edge of the ionized flow (upper right corner).  The emission peaks at two additional locations in the map: first, at the convergence of W1/W2/W3, and second at a point just inside the edge of the flow of lower ionization spectral lines.  Top Right: The same region of sky mapped in [Fe II] 26 $\mu$m for comparison.  Bottom Left: The same plot for [Ne V] emission.  Bottom Right: [Fe II] 24.5 $\mu$m.  Although the [Ne V] detection is weak, the spatial distribution of the emission suggests that it is real, and spatially distinct from the stronger [Fe II] emission.}
\label{o4map}
\end{center}
\end{figure}

\begin{figure}
\begin{center}
\includegraphics[width=0.45\textwidth]{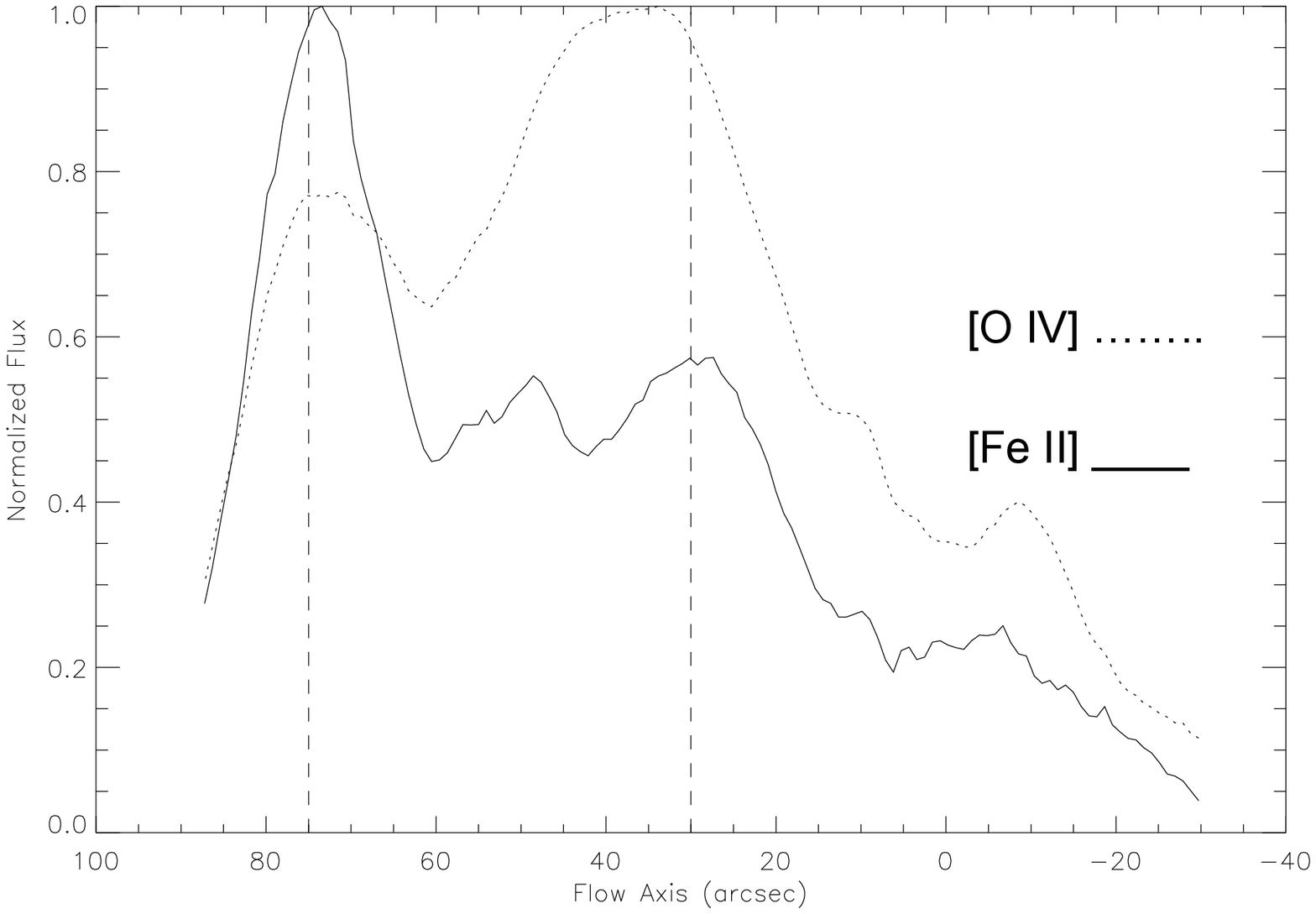}
\caption{Comparison of the peak of bow shock emission in a high excitation line, [O IV], and a lower excitation line, [Fe II], in terms of normalized flux (vertical axis), as a function of position on the flow axis (rotated 40 degrees), similar to the analysis in \citet{raines00}.  The solid curve represents [Fe II].  The dashed line represents [O IV].  The two vertical lines line up with the peak emission in[O IV], at the W2 and Jethead apertures.  Note that the ``Jethead'' peak is offset from the corresponding [Fe II] peak by $\sim$ 5'', or one full resolution element.}
\label{shockfront}
\end{center}
\end{figure}

The [Ne V] emission is confined to the area around W2 and a point at the northwest end of the flow (Figure \ref{o4map}), roughly coinciding with the edge of the lower excitation line emission from [Fe II].     \citet[][his figure 3.15]{raines00} noted that the peak of the [Fe II] (1.644 $\mu$m) emission was shifted to the east by $\sim$ 3'' from the peak of the H$_2$ (2.122 $\mu$m) emission; we note a similar result in our maps (Figure \ref{shockfront}) -- although the spatial resolution is considerably lower than in Raines' images -- and we can clearly see that the peak emission from all of the ionized species is located east of the peak of the molecular emission.  Additionally, the higher ionization species appear to be themselves shifted eastward by a further $\sim$ 5'' compared to the lower ionization species, suggesting that we are resolving the postshock region.

Many other species, including [Ne II], [Si II], and [S I] were detected in GGD37 in \citet{wright96} with the Infrared Space Observatory (ISO); however strong lines in the IRS data such as [Ne III] were {\it not} noted in the ISO data; this may be due to the large beam size of ISO and the small spatial extent of the [Ne III] emission.

\begin{deluxetable*}{l c r r r}
\tabletypesize{\scriptsize}
\tablecaption{Surface Brightness by Chemical Species by Aperture  \label{fluxtable}}
\tablehead{
\colhead{Species} & \colhead{Wavelength ($\mu$m)} &\colhead{Module} & \colhead{W123 Intensity} & \colhead{Jethead Intensity}}
\startdata
$[$O IV] & 25.9 & LH &  5.80 & 1.60 \\
$[$Ne V] & 24.3 & LH &  0.57 & 0.34 \\
$[$Ne III] & 15.5 & SH &  78.9 & 28.0 \\
$[$Ar III] & 21.8 & LH &  0.41 & -- \\
\enddata
\tablecomments{Surface brightness in units of 10$^{-6}$ erg/cm$^2$/s/sr of all species at 15'' half-power beam width (HPBW) apertures of three regions of interest: the W1-W2-W3 emission complex (W123), the extreme NW end of the GGD 37 flow (Jethead), and a selected region devoid of structure in the southeastern corner of the map (Continuum Region); see Figure \ref{apertures}.  In order to integrate the different modules, we utilized the flux of the [Ne II] and the continuum level as a normalization constant; the SH fluxes are thus decreased by a factor of $\sim$ 4.5 from observed values, and the SL2 spectra are decreased by 1.33 to match continuum levels with SL1.  A  `` -- '' marking indicates a non-detection or failed linefit.}
\end{deluxetable*}


\section{Discussion}

Originally detected as a radio emitter by \citet{hughes82} and resolved into three separate components (W1, W2, and W3) by \citet{hughes89} using VLA data, W2 is a prime candidate for an intermediate mass protostellar object driving an exceptionally powerful outflow \citep{garay96}.  The edge of the GGD 37 flow is approximately located at Source S as defined in \citet{hartigan85}, and the W2 and S locations coincide with a line of soft X-ray emission \citep{pravdo05}, as well as velocity-resolved [O III] (5007 \AA) emission in the case of Source S \citep{hartigan86}.  \citet{pravdo05} identify W2 and S as separate exciting sources for soft X-rays, although they observe significant emission from the line connecting the two as well, and the [O III] emission has linewidths of $\sim$ 420 km s$^{-1}$.  It is therefore not very surprising that such highly ionized species as O$^{3+}$ and Ne$^{4+}$ are present; these ions trace the hottest gas and provide constraints on the shock energetics.

Our data support the identification of GGD 37 as a protostellar outflow originating at W2. The continuum spectral index $\alpha$ is defined as $d~log(\lambda~ F_{\lambda})/~d~log(\lambda)$.  Calculated at the position of the W1-2-3 aperture (Table \ref{fluxtable})  $\alpha$ $\sim$ 1 when measured from 5.5 to 10 $\mu$m and $\sim$ 3 when measured from 10 to 35 $\mu$m, both results in the range typical of Class 0/I YSOs.  Our data reveal spectral lines with extremely high excitation potentials, ranging up to 97 eV in the case of [Ne V].  The ratio of [Ne II]/[Fe II] (26 $\mu$m) which peaks in the vicinity of W2, is 10 - 20 -- much larger than is seen in other intermediate mass YSO regions \citep[e.g. HH 54 or HH 7-11][]{neufeld06a}.  The [Ne III]/[Ne II] ratio is $\sim$ 0.2-0.5 in this region, gradually falling off to the west before peaking slightly again at the edge of the flow, although it should be noted that this may be confused by separate excitation mechanisms.  Ne$^+$ ions can be produced in T Tau disks by X-ray absorption in the warm disk, and the recombination of Ne ions is slow.  Ne$^{2+}$ has only weak charge transfer, and Ne$^+$ none at all, with H \citep{glassgold07}.  This is of importance in analyzing [Ne II] in protoplanetary disks.  In the present case the [Ne III] and [Ne II] emission regions are so similar in shape, they are likely to be excited the same way, in an extended and unusually powerful protostellar outflow.  The spectra closely resemble that of a supernova remnant, although there are several differences: the lack of certain lines such as [P II] 32.5 $\mu$m and [Ne III] at 36 $\mu$m \citep{neufeld07}, although the latter is probably too faint to be detected in this case.  Additionally, [O IV] and [Ne V] are not observed in HII regions.

Published models \citep[e.g.][hereafter HM89]{hollenbach89} don't seem to account for the observed line ratios, however.  The first problem comes in measuring the mechanical luminosity of the flow from cooling rates.  [O I] 63 $\mu$m is modeled as the principal cooling line, and [Fe II] and [Si II] are both expected to trace [O I].  Predicted intensity ratios of [Fe II]/[O I] and [Si II]/[O I] are $\sim$ 0.1 for most values of electron density (n$_e$) and shock speed (v$_s$), as these lines are also optically thin.  [O I] intensity is proportional to the total mass flow through the shock boundary; using the ``W123'' aperture we calculated an expected intensity of $\sim$ 2 $\times$ 10$^{-4}$ erg s$^{-1}$ cm$^{-2}$ sr$^{-1}$, and the resulting mass flow rate is 10$^{9}$ n$_0$ v$_s$ for density n$_0$.  Assuming a fast shock velocity of 100 km s$^{-1}$, and n$_0$ $\sim$ 2 $\times$ 10$^4$ cm$^{-3}$, the mass flow rate is 3.3 $\times$ 10$^{-6}$ $\msunyr$.  If instead we perform the same calculation with [Si II] (34.8 $\mu$m), the mass flow is 5.4 $\times$ 10$^{-5}$ $\msunyr$, nearly 10 times higher.  This is due to the ratio of [Si II]/[Fe II] flux $>$ 1 over the entire flow, which does not match any conditions in HM89.  Although the [Si II] and [Fe II] emission are morphologically quite similar, this may reflect differing critical densities, or a high Si/Fe abundance compared to solar values.  However this is more likely because HM89 do not include sufficiently fast outflows.

Additionally HM89 do not predict fluxes of high-excitation species such as [Ne V] and [O IV].  The Mappings III Shock Model Library \citep{allen08} predicts mid-IR flux ratios, as already noted for the case of [Ar III].  The [Ne II]/[O IV] ratio is $\sim$ 500 in the W2 region and 190 at the leading edge of the flow.  The ratio of [Fe II] (26 $\mu$m) to [O IV] is 14.3 and 41.8, respectively.  [Ne III]/[Ne II] is $\sim$ 0.21 and 0.15, respectively.  [S IV] is not detected, but an upper bound to [S IV]/[S III] is $\sim$ 0.11 and 0.35.  The relative lack of [O IV] compared to [Ne II] and [Fe II] predicts a large shock velocity of 500 km/s or more,  without a precursor.  The lack of [S IV] and relatively high [Ne III] flux seem to indicate a slower 200 km s$^{-1}$ shock, but still without a precursor.
Mappings III is far too low in density (and therefore in recombination rates), and does not fit the observed line ratios.

Modeling the molecular emission and the forbidden lines simultaneously is difficult.   \citet{flower03} model GGD 37 itself with a 25 km s$^{-1}$ J-shock with a magnetic precursor, which matches the H$_2$ emission well but is not nearly fast enough to produce the high-excitation lines.  In advance of detailed complex modeling we can make cruder estimates of shock energetics implied by the mid-IR and X-ray emission.

\begin{figure*}
\begin{center}
\includegraphics[scale=0.7]{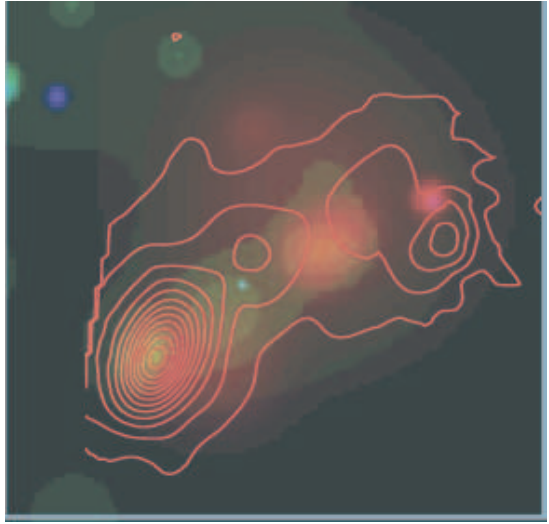}
\caption{Color contour map of 0.3-1.5 keV X-ray emission of a 1' $\times$ 0.8' region in GGD37; overplotted in red unfilled contours is the [O IV] Spitzer-IRS emission, overlaid in color intensity contours is the [O IV] emission from W2 (lower left corner) to the edge of the ionized flow (upper right corner).}
\label{xray_ir}
\end{center}
\end{figure*}

We can estimate the shock speed from detailed balance from excitation by UV/X-ray emission, as a function of the ratio of [Ne V]/[O IV].  The highest temperature achieved in the shock is given by the [Ne V] 97 eV transition, indicating a temperature of 4 $\times$ 10$^6$ K.  We can calculate the shock speed from the observed temperature and luminosity \citep[e.g.][]{watson07a}.  Assuming a solar abundance ratio of O/Ne $\sim$ 8, we derive a post-shock temperature of 18,100 K for the observed [O IV]/[Ne V] ratio of $\sim$ 10.  This indicates a post-shock velocity of only 24 km/s.  Furthermore, \citet{schneider09a} posit that the entire flow is one continuous flow of shock-heated material with a velocity in the hundreds of km s$^{-1}$.  The approximate emitting volume is $\sim$ 4 $\times$ 10$^{52}$ cm$^3$, and the derived n$_e$ of emitting plasma is $\sim$ 10 cm$^{-3}$ in that region.  Thus we find a total shocked mass 0.62 $\msun$ for a density of 2 $\times$ 10$^4$ cm$^{-3}$.  The total momentum of the flow is then 124 $\msun$ km s$^{-1}$, assuming a fast shock speed of at least 200 km s$^{-1}$.  This is quite high compared to other powerful flows observed with Spitzer-IRS \citep[e.g. 1-2 $\msun$ km s$^{-1}$ for HH 7-11;][]{maret09}, further evidence that GGD 37 is unusual.  

If we propose a single emitting region containing all three observed ionization states of Ne (Ne$^+$, Ne$^{2+}$, Ne$^{4+}$), supposing the hot postshock gas is uniform in n$_e$ and temperature T, and the ionization states are in equilibrium, we could in theory estimate the physical conditions.  However under these assumptions it is impossible for a single region to produce all three lines simultaneously.  Thus comparable intensities of all three transitions suggests either a photoionizing source or a collisionally ionizing source -- in either case, likely a powerful and extremely hot jet, highlighted by the X-ray emission region (Figure \ref{xray_ir}).

A detailed discussion of the H$_2$ emission, as well as the [Fe II] and other fine structure line maps will be explored in subsequent papers.

\section{Conclusions}

We present the detection of [Ar III], [O IV], and [Ne V] fine structure lines, the highest appearance potential for optical or infrared lines ever detected in the vicinity of a Herbig-Haro flow.  Considering this data along with 0.2-1.0 keV X-ray maps of the flow, we suggest that GGD 37 is the hottest HH object detected so far.

The extended nature of the high ionization energy lines and soft X-ray emission from the GGD 37 flow suggests that it is extraordinarily energetic for a Herbig-Haro object, perhaps an indirect detection of a radiative jet.  In agreement with models of planar C+J shocks, we detect a significantly extended transitional boundary layer.  This suggests that the radio source W2 is a plausible candidate for a high mass embedded protostar driving an outflow to the northwest.

\acknowledgements

The authors would like to thank Neal Evans, Mike Dunham, Manoj Puravankara, Dave Pooley, and Peter Yoachim for helpful comments and discussion, and Klaus Hodapp for the use of K' images of Cep A. This work is based on observations made with the {\it Spitzer Space Telescope}, which
is operated by the Jet Propulsion Laboratory, under NASA contract 1407.  Support for
this work was provided by NASA through contract 1257184.
This research was supported in part by
Jet Propulsion Laboratory (JPL) contract 960803 to Cornell University,
and Cornell subcontracts 31419-5714 to the University of Rochester.  
This research has also made use of the SIMBAD database,
operated at CDS, Strasbourg, France.


\end{document}